# Uncertainty, incompleteness, chance, and design

Fernando Sols
Departamento de Física de Materiales
Universidad Complutense de Madrid
E-28040 Madrid, Spain

The 20th century has revealed two important limitations of scientific knowledge. On the one hand, the combination of Poincaré's nonlinear dynamics and Heisenberg's uncertainty principle leads to a world picture where physical reality is, in many respects, intrinsically undetermined. On the other hand, Gödel's incompleteness theorems reveal us the existence of mathematical truths that cannot be demonstrated. More recently, Chaitin has proved that, from the incompleteness theorems, it follows that the random character of a given mathematical sequence cannot be proved in general (it is 'undecidable'). I reflect here on the consequences derived from the indeterminacy of the future and the undecidability of randomness, concluding that the question of the presence or absence of finality in nature is fundamentally outside the scope of the scientific method[1].

**Introduction**

Since the publication in 1859 of Charles Darwin's classic work *On the origin of species*, there has been an important debate about the presence or absence of design in nature. During the 20th century, progress in cosmology has permitted to extend that debate to include the history of the universe. The intellectual discussion has become especially intense in the last few decades after the so-called "intelligent design" has been proposed as a possible scientific program that would aim at proving the existence of finality in biological evolution[2]. In this often unnecessarily bitter controversy, chance plus natural selection on the one hand and intelligent design on the other hand, compete as possible driving mechanisms behind the progress of species. Chance, or randomness, is no doubt an essential concept in many fields of science, including not only evolution biology but also quantum physics and statistical physics. It is surprising however that, within the mentioned controversy, it has been barely noticed that, within mathematics, randomness cannot be proved. More precisely, Gregory Chaitin has proved that the random character of a mathematical sequence is, in general, undecidable, in the sense proposed by the mathematicians Kurt Gödel and Alan Turing[3].

---

[1] This chapter is based on the longer article by F. Sols, *Heisenberg, Gödel y la cuestión de la finalidad en la ciencia*, in *Ciencia y Religión en el siglo XXI: recuperar el diálogo*, Emilio Chuvieco and Denis Alexander, eds., Editorial Centro de Estudios Ramón Areces (Madrid, 2012).

[2] W. A. Dembski, *Intelligent Design: the Bridge between Science and Theology* (1999).

In this chapter we shall argue that Chaitin's work, combined with our current knowledge of quantum physics, leads us inevitably to the conclusion that the debate on the presence or absence of finality in nature is fundamentally outside the scope of the scientific method, although it may have philosophical interest. The argumentation will take us to review some decisive moments in the history of physics, mathematics, and philosophy of science. We will refer to Newton's physics, Poincaré's nonlinear dynamics, Heisenberg's uncertainty principle, the wave function collapse, Gödel's theorems, Turing's halting problem, Chaitin's algorithmic information theory, Monod's philosophy of biology, the intelligent design proposal, and Popper's falsifiability criterion.

The guiding theme of our argumentation will be the attempt to address a fundamental question that is as simple to formulate as difficult to respond: "What or who determines the future?" We hope that the reflections presented here will contribute to clarify the debate on finality, helping to distinguish between established scientific knowledge and the philosophical interpretation of that knowledge.

**Practical indeterminacy in classical physics: Newton and Poincaré.**

In his monumental work *Philosophiae naturalis principia mathematica* (1687), Isaac Newton formulated the law of gravitation and the laws of classical mechanics that bear his name. The study of those laws with the tools of infinitesimal calculus, which he and Leibniz created, led to a deterministic world view in which the future of a dynamical system is completely determined by its initial conditions, more specifically, by the initial values of the positions and linear momenta of the particles involved together with the relevant force laws.

---

[3] G. Chaitin, *Randomness and Mathematical Proof*, Sci. Am. 232, 47 (1975).

This mechanistic picture of nature took strong roots, favored by the impressive success of Newton's mechanics in accounting for planet motion and ordinary life gravity. Despite not being confirmed by modern physics, determinism still enjoys some adherents.

In the late 19$^{th}$ century, Henry Poincaré addressed the three-body problem and concluded that the evolution of such a dynamical system is in general chaotic, in the sense that small variations in the initial conditions lead to widely different trajectories at long times. The longer the time interval over which one wishes to predict the system's behavior with a given precision, the more accurate the knowledge of the initial conditions must be. Poincaré concluded that the appealing regularity of the two-body problem, exemplified by the motion of planets around the sun, is the exception rather than the rule. The vast majority of dynamical systems are chaotic, which means that the prediction of their long time behavior is, in practice, impossible. Thus we encounter practical indeterminacy in classical physics.

One could still claim that, even if ruled out for practical reasons, determinism could still survive as a fundamental concept. One could argue that the future of nature and the universe, including ourselves, are determined indeed but in such a way that in practice we can make reliable predictions only in the simplest cases. For practical purposes, this determinism could not be distinguished from the indeterminism in which we think we live. Below we argue that quantum mechanics rules out a deterministic world view on not only practical but also fundamental grounds.

**Intrinsic indeterminacy in quantum physics: Heisenberg.**

During the first third of the 20th century, quantum mechanics was discovered and formulated. For the present discussion, we focus on a particular aspect of quantum mechanics: Heisenberg's uncertainty principle. It tells us that, due to its wave nature, a quantum particle cannot have its position and momentum simultaneously well defined. More specifically, if Δx and Δp are the uncertainties in position and linear momentum, respectively, then the following inequality must always hold:

$\Delta x \, \Delta p \geq h / 4\pi$     (1)

where *h* is Planck's constant. If we combine Poincaré's nonlinear dynamics with Heisenberg's uncertainty principle, we conclude that, in order to predict an increasingly far future, a point will be reached where the required knowledge of the initial conditions must violate the uncertainty principle. The reason is that the condition Δx→0 and Δp→0 (necessary to predict the far future) is incompatible with the inequality (1). We thus reach the conclusion that, within the world view offered by modern quantum physics, the prediction of the far future is impossible, not only in a practical but also in a fundamental sense: the information on the long term behavior of a chaotic system is nowhere [4]. Noting that non-chaotic systems are rare and always an approximation to reality, we may conclude that the future is open[5].

A remarkable example is Hyperion, an elongated moon of Saturn with an average diameter of 300 km and a mass of $6 \times 10^{18}$ kg whose rotation is chaotic. Zurek has estimated that quantum mechanics forbids predictions on its rotation for times longer than 20 years[6].

The deterministic view is still defended by the supporters of hidden variable theories. They claim that, below the apparent indeterminacy of quantum systems, there are some 'hidden variables' whose supposedly precise value would entirely determine the future. The experiments by Alain Aspect in the early 80s, as well as the numerous experiments on quantum information performed ever since, all inspired by the seminal work of John Bell, have ruled out the very important class of local hidden variable theories.

**Uncertainty vs. indeterminacy**

The wave nature of quantum particles is reflected in the superposition principle, which states that, if A and B are possible states of a system, then a linear superposition of them such as e.g. C = A + B represents another possible state of that system (with some notable exceptions that will not be discussed here). On the other hand, a state A is said to be an eigenstate of an observable (physically measurable quantity) S if this observable has a well-defined value (without uncertainty) in A, for instance $s_A$. Similarly, let us assume that B is another eigenstate of S with eigenvalue $s_B$ different from $s_A$. We say that the physical quantity S has no uncertainty in the states A or B taken separately. However, S is uncertain in the state C because it is a linear combination of two eigenstates of S (A and B) with different eigenvalues ($s_A$ and $s_B$).

---

[4] Rolf Landauer used to say that "information is physical". Without a physical support, there is no information. The latter emerges as the various possible future evolutions become specified.

[5] K. Popper, *The Open Universe: An Argument for Indeterminism* (1982).

[6] W. H. Zurek, *Decoherence, Chaos, Quantum-Classical Correspondence, and the Algorithmic Arrow of Time.* Physica Scripta T76, 186 (1998).

In Schrödinger's wave mechanics, the state of a system evolves linearly in a deterministic way. So we can say that the uncertainty about S in the state C evolves deterministically. However, when we perform an ideal measurement of S, i.e., if we "ask" the system (prepared in state C) which value holds for the physical quantity S, then the "answer" (outcome of the measurement) can only cast one eigenvalue of S ($s_A$ or $s_B$). Quantum mechanics can only predict the statistics of many measurements of the observable S performed on the system always prepared in state C. In a single experiment, we obtain either $s_A$ or $s_B$. Then the measurement postulate (which must be added to Schrödinger's wave mechanics) tells us that, starting in state C before the measurement, the system is projected onto state A or B depending on whether the result of the measurement is $s_A$ or $s_B$. This process is referred to as the "collapse of the wave function". So because of the initial uncertainty in the value of S, the final state of the system after one measurement is undetermined.

In summary, the indeterminacy of the future results from the combined effect of the initial uncertainty about an observable in a given state and the collapse of the wave function when a measurement of that observable is made on the system. The evolution of a chaotic system coupled to an environment can be roughly viewed as a succession of quantum measurements, each one with a random outcome.

**What or who determines the future?**

We have already noted that quantum physics offers a world view in which the future is not entirely determined. This indeterminacy is not practical but fundamental. It permits us to think that our experience of free will may be real instead of merely subjective. If the outcome of a quantum process is undetermined, there is no fundamental reason to deny that possibility in some neural events which may be a macroscopic amplification of microscopic processes where quantum indeterminacy plays an essential role[7].

---

[7] The neurobiologist John C. Eccles (1963 Nobel Prize in Medicine) identified a neural process that might underlie an act of free choice [J. C. Eccles, *Evolution of consciousness*, Proc. Natl.

We proceed with our attempt to answer the question which gives title to this section: "What or who determines the future?" We have ruled out the determinism of the initial conditions. We may invoke some type of design (or finality) that would condition the evolution of a system to the achievement of a previously desired goal.

There is a type of design, which we shall label here as 'internal', that is not particularly controversial. Except for recalcitrant determinists, everybody agrees that there are engineers who design cars and architects who conceive buildings. Even a tidy teenager's room suggests the action of a designer who has ordered it. However, the consensus about internal design does not settle the question because many natural phenomena are not directly induced by a human being.

There is another type of design, which may be called 'external', that is polemic because it suggests the idea of transcendence. If internal design reflects the free action of human beings, external design would reflect what in the theological language would be called the action of divine providence, by which we mean the influence of God on the world without manifestly altering its laws. The picture of an undetermined world leaves room for –but of course does not prove– the existence of freedom and providence. Free will may act through *a priori* undetermined quantum processes which are likely to take place in our brain (see previous section).

The possible physical support for the action of providence is more difficult to delimit, probably because it is more general. However, the ubiquity of chaotic macroscopic systems, ranging from meteorology to the history of the solar system and the fluctuations of the primitive universe, strongly suggests that the long term indeterminacy of those systems is ultimately of a quantum nature and thus intrinsically open to a variety of evolutions. Returning to the example of Hyperion, if we try to predict the detailed rotation of that Saturn satellite in a century from now, we quickly realize that a wealth of drastically different evolutions is possible, all being compatible with the laws of physics and with the most detailed knowledge we may have of its present motion state. As noted before, the reason is that a detailed prediction of

---

Acad. Sci. USA **89**, 7320 (1992); F. Beck, J. C. Eccles, *Quantum aspects of brain activity and the role of consciousness*, Proc. Natl. Acad. Sci. USA **89**, 11357 (1992)]. Other neuroscientists question the reality of objectively free choice [K. Koch, K. Hepp, *Quantum mechanics in the brain*, Nature 440, 611 (2006); K. Smith, *Taking aim at free will*, Nature 477, 23 (2011)]. In regard to the work by Koch and Hepp, we wish to point out that there is much more in quantum mechanics than just quantum computation.

Hyperion's rotation in one century from now would require so precise a knowledge of its present state of motion that, even for a system of $10^{18}$ kg, it would have to violate Heisenberg's uncertainty principle. The information on the long term rotation of Hyperion is nowhere because it has no possible physical support.

Within the context of evolution biology, the concept of 'intelligent design' has been proposed in the last few decades. Its existence would be necessary to explain the emergence of complex biological structures that would be very unlikely *a priori*. Intelligent design is not very different from the external design we have mentioned above. The problem with the intelligent design program is that it is presented as a scientific program despite the fact that, as will be argued later, the questions about finality in nature lie outside the scope of the scientific method. Intelligent design may be an interesting philosophical or theological proposal, but not a scientific one. We will argue that the same can be said about the absence of design.

To explain the apparently unpredictable behavior of complex systems, especially in the context of biological evolution, the concept of *chance* (or *randomness*) is often invoked. Chance can be understood here as indeterminacy without design. Randomness is a ubiquitous concept not only in evolution theory but also in quantum and statistical physics. Quantum mechanics is successful in predicting the statistical outcome of experiments provided that these are performed on identically prepared systems. This requirement cannot be satisfied in biological history or in our personal lives. Quantum mechanics has little predictive power on experiments of uncertain outcome that can be run only once.

Thus, randomness is a useful concept when one studies the statistical behavior of many processes each of which carries some indeterminacy. The problem with randomness is that, as we shall see, it can never be confidently ascribed to any sequence of properly quantified events. The reason is fundamental because it is a consequence of Gödel's theorems, perhaps the most important result in the history of knowledge.

In the coming sections we try to understand the meaning and the epistemological consequences of the impossibility of proving randomness.

**Gödel's theorems**

In the 1920s, David Hilbert proposed a research program whose goal would be to prove, for (axiomatic) arithmetic and set theory, the following statements: (i) It should be possible to prove the consistency of the axioms, i.e., that they do not lead to contradictions. (ii) It should be possible to prove their completeness, i.e., that all theorems are derivable from the axioms. (iii) It should be possible to prove their decidability, i.e., that there is a general algorithm such that, when applied to any meaningful formula, would stop at some moment with only two possible outcomes, "yes" if the formula can be inferred from the axioms and "no" if it cannot.

The young Austrian mathematician Kurt Gödel tackled the problem and found that Hilbert's expectations where fulfilled for first-order logic. However, when he confronted the case of axiomatic arithmetic and set theory, he found a negative result, to the surprise and deception of many. Specifically, for Peano arithmetic Gödel proved: (i) It is incomplete, i.e., there is at least one formula such that neither the formula itself nor its negation can be derived from the axioms. (ii) The consistency of the theory cannot be derived from the axioms of the theory. (iii) The theory is undecidable, i.e., there is no general algorithm to decide whether a meaningful formula is derivable or not from the axioms of the theory.

These surprising negative results demolished Hilbert's expectations, according to which mathematics would be akin to a mechanical game where, from a finite set of axioms and logical rules, and with sufficient patience and skill (or with the help of a modern computer) all theorems could eventually be proved, and where an algorithm would exist which would tell us whether a meaningful formula is derivable form the axioms. The dream of mathematical certainty and of the systematic exploration of mathematical truths was vanishing forever. Our conception of mathematics becomes not so different from our conception of a physical theory. If a physical theory postulates universal laws with the hope that no experiments will be found that refute it, mathematics postulates a set of axioms with the hope that they will not lead to contradictions.

The English mathematician Alan Turing put Gödel's undecidability on a more tangible ground. Well before computers had been invented, he conceived a mathematical program as a sequence of zeros and ones such that, when fed with an input (itself a sequence of 0 and 1), it would yield an output that would also be expressed as a series of 0 and 1. This ideal concept of universal computer is referred to as Turing's machine.

Turing proved that no algorithm exists such that, when applied to an arbitrary self-contained program (which includes the input), always stops and yields one of two possible answers: 1 if the self-contained program stops, and 0 if it does not. That is, Turing proved that the question of whether or not a program goes to a halt is undecidable. In short, 'the halting problem' is undecidable.

Since the work of Turing, other classes of problems have been identified as undecidable. In some cases, the undecidability is suspected but has not been proved. An important example of undecidable problem is that of asserting the random character of a mathematical sequence.

**Randomness**

Chance is a concept that, in a vague form, has been invoked since remote times. It is used by the ancient Greeks, especially Aristotle, and even in the Bible. Chance can be understood as indeterminacy without design. Randomness is practically a synonym of chance. The term 'chance' has a dynamic connotation and is used more often in the context of history and biology; 'randomness' has a more static meaning and is preferentially used in physics and mathematics. For many practical purposes, the two terms can be taken as synonymous, since the inability to predict the future is directly related to the absence of a clear pattern in the (conveniently quantified) past events. However, they are not always equivalent. For instance, as a result of chance, it is possible, with a low probability, to generate a non-random sequence[8].

The Argentine-American mathematician Gregory Chaitin has given a static but probably more fundamental definition of randomness that applies to mathematical sequences[9], which is not an important limitation if one is aiming at a quantitative description of nature. It is simpler to define the absence of randomness. A mathematical sequence (made, for instance, of zeros and ones) is not random if it can be compressed, i.e., if there is a shorter sequence that, when applied to a Turing machine, yields the longer sequence. Then one says that the shorter sequence contains the longer sequence in a compressed form. A long sequence is said to be random if it cannot be compressed, i.e. if no shorter sequence exists that determines it.

---

[8] For a detailed discussion, see [A. Eagle, *Randomness is unpredictability*. British Journal for the Philosophy of Science, 56 (4), 749–790 (2005)] and [A. Eagle, "Chance versus Randomness", *The Stanford Encyclopedia of Philosophy (Spring 2012 Edition)*, Edward N. Zalta (ed.), URL = http://plato.stanford.edu/archives/spr2012/entries/chance-randomness/].

[9] G. Chaitin, *Meta Math! The Quest for Omega.* Vintage Books (New York, 2005).

A canonical example of non-random sequence is, for instance, the first million digits of the number π (up to $10^{13}$ digits of π are known). Despite its apparent random character, that sequence is not random because a program can be written, of length much shorter than one million digits, which yields number π as the outcome.

Chaitin has proved that the question of whether or not a long number sequence is random, is undecidable, in the sense of Gödel and Turing. No algorithm exists that, when applied to an arbitrary sequence, goes to a halt casting an answer 'yes' or 'no' to the question of whether that sequence is random.

The consequence is that, even if chance or randomness is a useful –even necessary– hypothesis in many contexts, it cannot be ascribed with total certainty to any mathematical sequence and therefore to any physical or biological process. This consideration may not have practical implications, but it has indeed important epistemological consequences. To the extent that chance is understood as indeterminacy without design, it can never be legitimate to present the absence of design as a scientific conclusion. Randomness can be a reasonable working hypothesis, a defensible philosophical proposal, but it cannot be presented as an established scientific fact when questions of principle are being debated such as the presence or absence of design in nature.

**Popper's falsifiability criterion**

In his work *Logik der Forschung* (1934), the philosopher of science Karl Popper proposed that the demarcation line to distinguish genuinely scientific theories from those which are not, is the possibility of being *falsifiable*, that is, the possibility of conceiving an experiment among whose possible results there is *a priori* at least one that would contradict the prediction of the theory. This means that a theory containing universal statements cannot be verified (as an infinite number of verifying experiments would have to be made) but can be falsified (as a single, properly confirmed experiment is sufficient to refute the theory). Within this picture, all scientific theories are provisional in principle. However, when a theory correctly explains and predicts thousands of experimental facts after decades of collective scientific work, we can practically view it as a correct description of nature. This is the case e.g. of atomic and quantum theories. They started as bold proposals in the early 19[th] and 20[th] centuries, respectively, and nowadays we are as certain about them as we are about the spherical shape of the Earth.

The universal statements that compose a scientific theory are proposed from the empirical verification of many particular (or singular) statements, following an inductive process. Those singular statements must express legitimate certainties, in what refers both to the validity of the mathematical language employed and to the claim that they correctly describe reality. If the observation is made that planets follow elliptic trajectories around the sun, one is implicitly assuming that, within a certain level of approximation, one is entitled to associate a set of empirical points with the mathematical concept of ellipse.

It is this latter requirement, namely, that a set of numbers can be associated with a mathematical object, what cannot be satisfied when a supposedly universal statement invokes randomness. As noted above, the reason is that randomness cannot be ascribed with certainty to any mathematical sequence. And here we cannot invoke the qualification 'within a certain level of approximation', since an apparently random sequence like the first million digits of π is in fact radically non-random.

This observation is compatible with the fact that, for many practical purposes, the first million digits of π can be taken as random. However the above described limitation is important when we refer to laws that invoke randomness with a pretension of universality, especially if the link to randomness is used to draw metaphysical conclusions (such as the absence of design in nature), and even more particularly if those philosophical proposals are presented as part of the established scientific knowledge. All this is reconcilable with the fact that randomness is a useful –even essential– hypothesis in many scientific contexts. However, chance is not a scientific datum that can be used to reach philosophical conclusions.

**Chance in the interpretation of evolution biology**

The scientific evidence in favor of the historical continuity and the genetic relationship among the diverse biological species is overwhelming, comparable to the confidence we have on the atomic theory[10]. However, for reasons we have already anticipated, the same cannot be said of an ingredient that is often included in the description of evolution biology. The problem is not methodological, since, as already noted, chance or randomness is a useful working hypothesis in many fields of science. The problem appears when the concept of chance is considered sufficiently established to take it to the domain of principles, where philosophical ideas are debated.

---

[10] F. J. Ayala, *Darwin and Intelligent Design* (2006).

In his influential work *Le hasard et la necessité (Essai sur la philosophie naturelle de la biologie moderne)* (1970), the French biologist and Nobel laureate Jacques Monod contrasts chance and natural selection as the two driving mechanisms of evolution. Chance is indeterministic; natural selection is deterministic. He identifies chance with indeterminism without a project, but never defines chance in a quantitative form, except for some occasional reference to its possible quantum origin, which does not solve the mathematical problem. The lack of a precise definition does not seem to deter Monod from invoking chance as an essential concept. He presents it as the only possible source of genetic mutations and of all novelty in the biosphere, and then states that chance is the only hypothesis compatible with experience. As I may seem to exaggerate, I reproduce a quote below. After describing some genetic mutations, Monod writes[11]:

"...[mutations] constitute the *only* possible source of modifications in the genetic text, itself the *sole* repository of the organism's hereditary structures, it necessarily follows that chance *alone* is at the source of every innovation, of all creation in the biosphere. Pure chance, absolutely free but blind, at the very root of the stupendous edifice of evolution: this central concept of modern biology is no longer one among other possible or even conceivable hypotheses. It is today the sole conceivable hypothesis, the only one that squares with observed and tested fact. And nothing warrants the supposition — or the hope — that on this score our position is likely ever to be revised."

One may understand that professor Monod was not aware of Chaitin's work on the fundamental non-demonstrability of randomness, which first appeared in the 1960s but seems not to have reached the biologists yet. However, it is more difficult to understand why he apparently ignored two ideas: (i) the old intuition (previous to Chaitin's work) that chance is, if not impossible, at least difficult to prove; and (ii) Popper's falsifiability criterion, knowing that it is difficult to conceive an experiment or observation that yields, as an unequivocal outcome, the absence of chance (at least,

---

[11] The emphases are by Monod.

Monod was not proposing one); that is, knowing that the chance hypothesis is non-refutable.

**Design and chance lie outside the scope of the scientific method**

While discussing the relation between Chaitin's work on randomness and the debate on finality, Hans-Christian Reichel made the following observation[12]:

"Is evolution of life *random* or is it based on some law? The only answer which mathematics is prepared to give hast just been indicated: the hypothesis of *randomness* is *unprovable* in principle, and conversely the *teleological* thesis is *irrefutable* in principle."

This logical conclusion may be viewed as a virtue of design theories, since they cannot be refuted. However, it may also be viewed as a weakness, since, according to Popper's criterion, a theory which is *fundamentally irrefutable* cannot be scientific.

We thus reach the conclusion that, due the impossibility of verifying randomness in any particular sequence of (conveniently quantified) events, finality cannot be refuted as a general law. This intuitive conclusion is rooted in Gödel's theorems.

Analogously, we may wonder whether design cannot be verified for a particular sequence of events, or equivalently, whether the chance hypothesis is irrefutable. To prove these two equivalent statements, we may seem to lack a fundamental theorem of the type invoked in the previous paragraph. However, the weakness of the chance assumption is not so much that it is *practically irrefutable* when invoked as an ingredient of a general law, but rather that it is *fundamentally unverifiable* when ascribed to any singular event properly characterized by a mathematical sequence. Popper's falsifiability criterion emphasizes that, in order to be considered scientific, a universal statement should be amenable to refutation by the hypothetical observation of a singular event that contradicted the general proposal. However, such a criterion gives for granted that the general law can at least be verified in a finite number of singular cases which provide the seed for induction. This latter requirement cannot be satisfied by the chance hypothesis, for fundamental reasons rooted in Gödel's theorems.

---

[12] H. C. Reichel (1997). *How can or should the recent developments in mathematics influence the philosophy of mathematics?*, in *Mathematical Undecidability, Quantum Nonlocality and the Question of the Existence of God*, A. Driessen and A. Suarez, eds., Kluwer Academic Publishers (Dordrecht, 1997). Emphases are by Reichel.

In brief, Chaitin's work on the undecidability of randomness leads us to conclude that the design hypothesis is irrefutable as a general law while the chance hypothesis is unverifiable in any particular case. Both types of assumptions lie therefore outside the reach of the scientific method.

A similar conclusion may be reached following some intuitive reasoning not necessarily rooted in fundamental theorems. To that end, we imagine a debate between two scientists who are also philosophers. Albert is in favor of chance; Beatrice favors design. They are shown two sequences describing two different natural processes. The first one seems random; the second one displays some clear non-random patterns.

They discuss the first sequence, unaware of Chaitin's work on Gödel and Turing. Albert claims that the sequence is obviously random, since it shows no clear pattern. Beatrice responds saying that the sequence is designed, although not manifestly so. She claims that the designer has wished to give the sequence an appearance of randomness. They don't reach an agreement.

Had they been informed about Chaitin's work, the debate would not have been very different. Albert would have continued to claim that the first sequence is random but admitting that he cannot prove it. Alice would have insisted on the ability of the designer to simulate randomness and quite pleased would have noted that the random character of the sequence was unprovable in any case. No agreement would have been reached.

Now they discuss the second sequence. Beatrice claims that, quite obviously, it has been designed, since it shows some clear patterns. Albert counters that the sequence is random, noting that, with a nonzero probability, a randomly generated sequence may happen to show some repetitive patterns. He goes on to point out that, given the physical content of the processes described by the second sequence, those patterns have been necessary for the existence of the two debaters. If the sequence had not shown those regularities, none of them would be there to argue about it. Thus, says Albert, the non-random character of the second sequence should not be surprising, as it is a necessary condition for the very existence of him and Beatrice. Again, they don't reach an agreement.

The two philosophical contenders are unable to reach an agreement and no experiment seems to exist that can settle the question. The debate we have just described is obviously a caricature of a real discussion. However, it is easy to find in it reasoning patterns that are frequently heard in debates about the presence or absence of design in natural processes, whether biological or cosmological. When there is a strong philosophical motivation to maintain an interpretation, there is always an argument to defend it in front of the experimental appearance. It is naturally so, because in the debate about finality no decisive experiment or observation can be conceived.

It is interesting to note that the existence of design is non-controversial in other contexts. Nobody questions the existence of design in an airplane, although strictly speaking it is not more provable or less refutable than in the case of biological evolution. There is no experiment that casts as a result that the airplane has been designed. The difference is that we have an experience of design in ordinary life; we know that there are engineers who design airplanes. However, we don't have similar evidence about the existence of an external designer promoting the evolution of species. Because of this, the question of design in biological evolution will always be more controversial.

We are led to conclude that the debate about the presence or absence of finality lies outside the scope of the scientific method. Returning to our two previous contenders, it is clear that, even if they are able to agree about the apparent random or non-random character of the sequences, the debater in a weak position always has an argument to deny the apparently winning interpretation. We have argued that the apparent irreducibility of the chance-design debate is actually fundamental, since it can be regarded as a consequence of Gödel's theorems.

It seems more constructive that, in their daily scientific work, the two researchers participating in the discussion just described, choose in each context the working hypothesis which best stimulates the progress of knowledge, leaving for the sphere of the philosophical interpretation those considerations about finality which can be debated with the tools of reason but not with the tools of the scientific method.

I would like to thank Gregory Chaitin, Javier Leach, Anthony Leggett, Miguel Angel Martín-Delgado, Javier Sánchez Cañizares, Ignacio Sols, Ivar Zapata and Wojciech Zurek, for discussions on the questions here addressed. Posthumously, I would also like to thank discussions with John Eccles and Rolf Landauer. This acknowledgement